\documentclass{elsarticle}
\usepackage{graphicx}
\usepackage{psfrag}
\usepackage{makeidx}  
\usepackage{multirow}
\usepackage{psfrag}
\usepackage{caption}
\usepackage{amsmath}

\usepackage{amsfonts}

\begin{document}

\title{Could the Hilbert Space Be a Smaller Place?\\A Neural Network Perspective}

\author[ca]{Jean~Michel~Sellier$^*$}
\address[ca]{MILA, Montr\'{e}al, Qu\'{e}bec, Canada\\$^*$\texttt{jeanmichel.sellier@gmail.com}}

\begin{abstract}
In quantum many-body problems, one of the main difficulties        comes from the
description of non-negligible interactions which require, at least in   principle,
an exponential amount of information.    Recently, in the context of spin glasses
and Boltzmann machines, it has been demonstrated that systematic machine learning
of the wave function   can   reduce    these issues  to a tractable computational
problem.         In this work, we apply this approach to a different    situation,
i.e. the problem of finding the ground state of a given    quantum system made of
electrons, entirely described by its Hamiltonian operator,       and by utilizing
feedforward neural networks.    Although still in the shape of a proof of concept,
one can already observe  that this seminal idea is able to substantially simplify
the complexity of this peculiar, and important, problem.
\end{abstract}


\maketitle

\section{Introduction}

The motto "Hilbert space is a big place" is well known by    the practitioners of quantum
mechanics. It describes, in a few words,       how the complexity of this space increases
rapidly when approached either theoretically or numerically. The purpose of this paper is
to show that neural networks can help to reduce such difficulties. Specifically, we focus
on the problem of finding the ground state of typical quantum systems such as one or more
electrons in the presence of a given external potential. This is twofold:   it shows that
single-electron problems can be efficiently dealt by neural networks,   and that  it also
can be applied to the context of quantum many-body problems    in both density functional
theory and first-principle approaches. This is a problem which is not only of theoretical
importance, but also of technological,   economical   and even sociological relevance. In
fact, quantum mechanics is the theory which has made possible  things such as laser beams,
integrated circuits, microprocessors,  and has even improved our comprehension of the DNA.

Recently,     Carleo et al.  introduced a very general method based on neural networks to
represent the state of a quantum system.  As a practical instance, this approach has been
applied to the specific case of Boltzmann machines representing the state of a spin glass
\cite{Carleo-2017}, \cite{Carleo-2018}.   Although at a seminal stage, this method proves
to be very robust and accurate.    Inspired by these results, we hereby take advantage of
the tenets of such approach and apply them  to the problem of finding the ground state of
a quantum system consisting of one, or more, electrons immersed  in an external potential.
To the best of the author's knowledge, such attempt has not been tried yet and, therefore,
the potential benefits of the application of feedforward neural networks in such physical
context remains substantially unknown.         Due to the impressive capacities of neural
networks to efficiently explore spaces with exponential complexity,         and therefore
represent very complex function mappings with relatively small resources,   \cite{Bishop},
\cite{Bengio} this might eventually allow one to tackle regimes  which have traditionally
been forbidden to other more standard numerical approaches. As a matter of fact,    it is
possible to report examples of systems in which both {\sl{stochastic}}                and
{\sl{deterministic}} approaches fail        (a good example being represented by the sign
problem in Quantum Monte Carlo methods).

More specifically, the method suggested in this paper is based on  representing      wave
functions by means of feedforward neural networks  (similar to \cite{Carleo-2017} but not
necessarily Boltzmann machines), and for different quantum systems  (i.e. not necessarily
spin glasses).  The network   is   then trained   by means of a minimization of the total
energy which is performed by a genetic algorithm.   Such representation is mathematically
guaranteed to behave properly due to the existence of the universal approximation theorem
\cite{Kolmogorov}, \cite{Cybenko}, \cite{Hornik}    (although the reader should note that
this theorem does not specify the amount of resources required for such task).   It turns
out that this idea, although quite simple,    seems to be effective and robust for a wide
range of quantum systems as we will show in this paper. In fact, comparisons between this
novel approach and known exact solutions, along with numerical solutions,     show a good
agreement,       although many aspects of this approach still need to be investigated and
understood.

This paper is organized as follows. First, we introduce the method in details. Afterwards,
it is applied        to a series of situations where the exact solution is known. Finally,
comparisons with    numerical solutions are presented as well.       To conclude, various
directions for further investigations to improve and better understand this seminal  idea
are discussed.

\section{The Ground State Problem}

Although,   nowadays,     many different formulations of quantum mechanics nowadays exist
(e.g. \cite{Schroedinger}, \cite{Keldysh}, \cite{Feynman}, \cite{Wigner}, \cite{SPF}), we
focus on the Schr\"{o}dinger formalism             which represents the standard approach.
Therefore, for the sake of self consistency,       we start by briefly recalling the main
tenets of this theory which is based on the concept of wave functions (for simplicity, in
the following exposition we limit ourselves to the non-relativistic case).       We, then,
present the problem of finding the ground state of a system in this approach.

\bigskip

{\sl{The time-independent Schr\"{o}dinger equation}}. The time-independent Schr\"{o}dinger
equation is an eigenproblem which eigenvalues represent     the allowed energy levels of a
system and which eigenfunctions  (or wave functions)     represent a complete mathematical
description of the physical system.   This equation, in the presence of a time-independent
external potential $V=V({\bf{x}})$, and for a particle with charge $-q$, reads:
\begin{equation}
 \hat{H} \psi \left( {\bf{x}} \right) = E \psi \left( {\bf{x}} \right),
\label{schroedinger_time_independent}
\end{equation}
where $\hat{H}$ is known as the Hamiltonian operator and reads:
\begin{equation}
 \hat{H} = \frac{\hat{p}^2}{2 m} -q V({\bf{x}}) = -\frac{\hbar^2}{2 m} \nabla_{{\bf{x}}}^2 -q V({\bf{x}}),
 \label{hamiltonian}
\end{equation}
with $m$ the mass of the particle and the operator
$\nabla_{{\bf{x}}}^2 = \frac{\partial^2}{\partial x^2} + \frac{\partial^2}{\partial y^2} + \frac{\partial^2}{\partial z^2}$.

The solution of the eigenproblem (\ref{schroedinger_time_independent})    can be formally
written as a set of ordered couples $(E_n, \psi_n)$ for $n=0,1,2, \dots$      (ordered by
increasing values of $E_n$), where the wave function          $\psi_n = \psi_n({\bf{x}})$
represents a stationary state,       i.e. not depending on time. Among all wave functions
obtained in this way,      the function $\psi_0 = \psi_0({\bf{x}})$ and its corresponding
energy $E_0$ (i.e. the minimum eigenvalue) have a special place in the theory and     are
known as    the {\sl{ground state}} wave function and energy respectively.    Finding the
ground state and energy of a given quantum system,          i.e. solving the eigenproblem
(\ref{schroedinger_time_independent}), represents the fundamental problem of this work.

\section{Neural Network Representation of Wave Functions}

A common practice in the numerical treatment of quantum systems consists in expressing  a
wave function   in terms of some given orthonormal basis (usually chosen according to the
problem at hand). Classically,   one expands the wave function in terms of a series which
reads:
\begin{equation}
 \psi(x) = \sum_{l=1}^N a_l \phi_l(x),
\label{series}
\end{equation}
for some arbitrarily fixed integer $N$, where the coefficients $a_l$ are complex  numbers,
and:
\begin{equation}
\int dx \phi_l^*(x) \phi_m(x) = \delta_{lm},
\end{equation}
for any integers $l$ and $m$ in the interval $[1, N]$   (with $\delta_{lm}$ the Kronecker
delta function).   In other words, this amounts to describe the wave function by means of
constants. Such a practice, on one hand, highly simplifies the computational problem  but,
on the other hand, it can introduce limitations in the representation.

%
The use of neural networks in this specific situation is relatively  new and is justified
by the universal approximation theorem  \cite{Kolmogorov}, \cite{Cybenko},  \cite{Hornik}.
In a few words,      this theorem states that a  feedforward network with only one single
hidden layer containing a finite number of non-linear units, or neurons,  can approximate
a given continuous function defined      on a compact subsets of the $n$-dimensional real
space, under mild assumptions on the activation function.      Therefore, a simple neural
network can represent a wide variety of interesting functions  when     given appropriate
parameters (i.e. weights and biases) and, in particular, it can represent a quantum state
or wave function.  However, the Reader should note that this theorem does not specify the
actual number   of      parameters needed to accurately approximate such functions.

\begin{figure}
\centering
\includegraphics[width=0.45\textwidth]{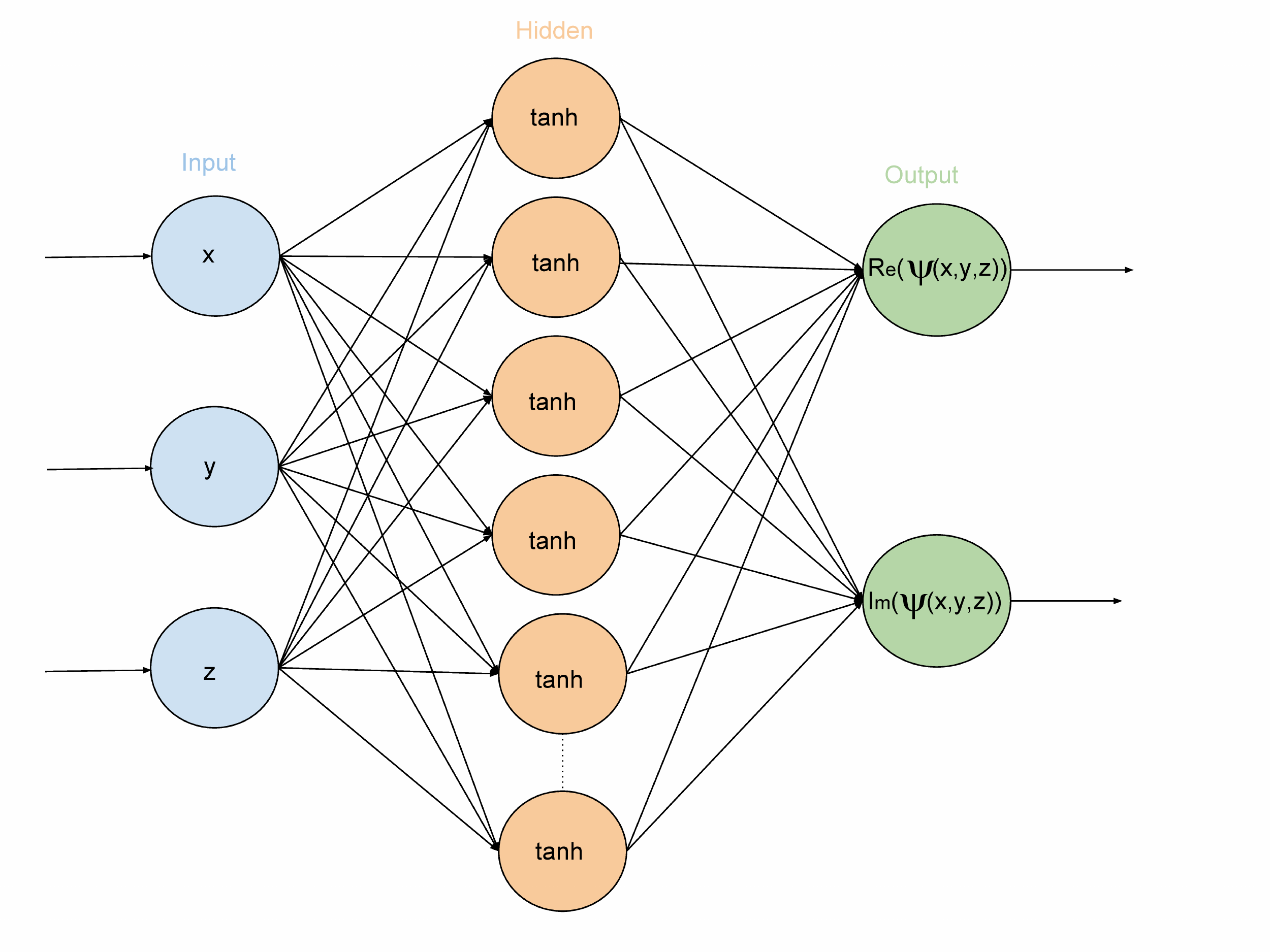}
\includegraphics[width=0.45\textwidth]{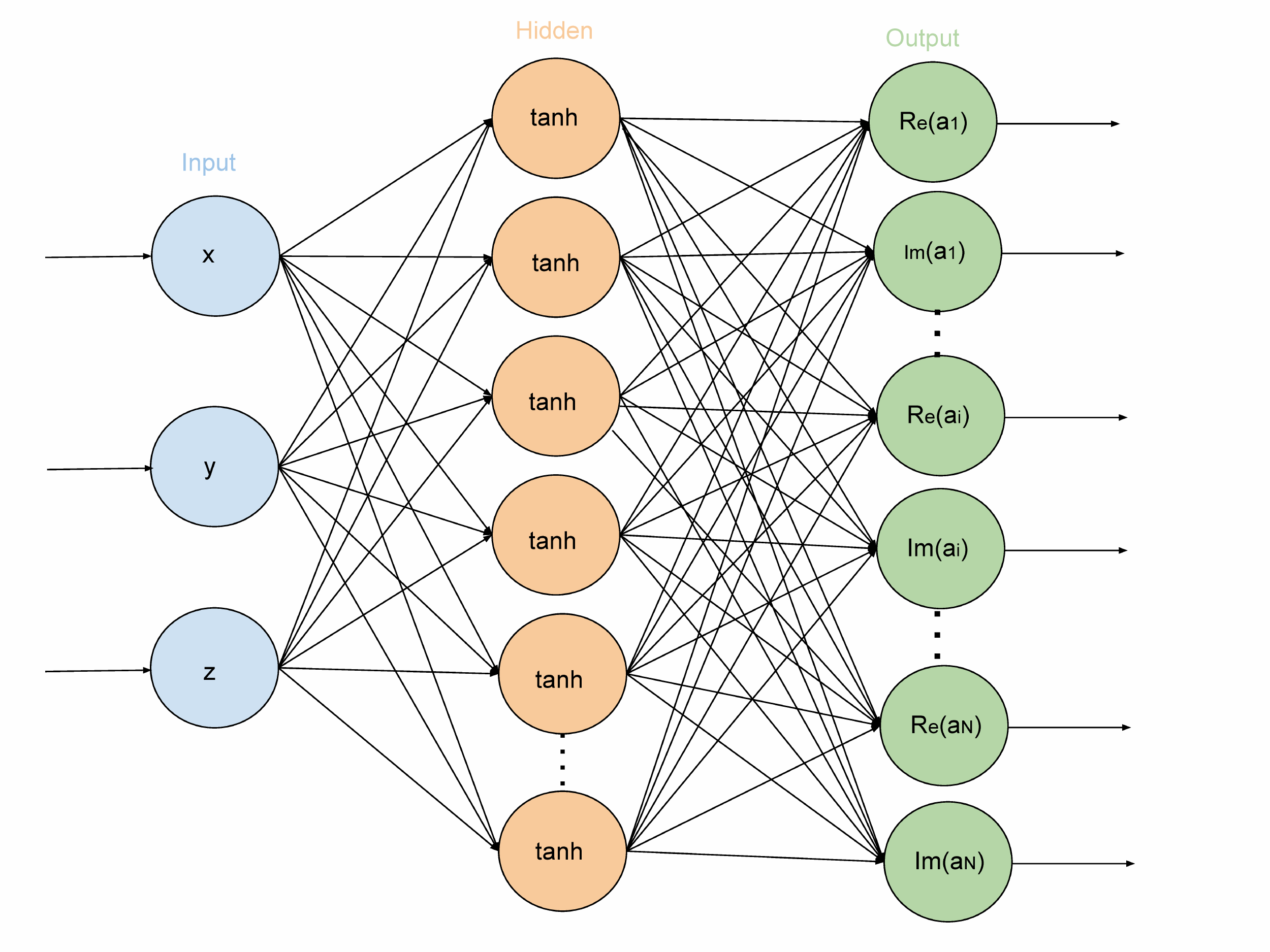}
\caption{Architectures of the networks used in this work. They both consist of an input layer which gets
the coordinates of the position, a non-linear hidden layer, and an output layer which in the first case (left-hand side) provides the
real and imaginary part of the probability amplitude, or wave function, and, in the second case(right-hand side) provides
the complex coefficients of the series (\ref{series}).}
\label{neural_network}
\end{figure}

Specifically,   in this work,   we utilize a simple feedforward neural network consisting
of an input layer,  one hidden layer with non-linear units and one linear output, and two
alternative approaches are depicted and utilized. In the first case,      the input layer
receives the position coordinates in the configuration space (i.e. $(x, y, z)$      for a
single-electron in a three-dimensional space)     and the output layer returns two scalar
values representing the real and complex part of the corresponding  probability amplitude
(see Fig. \ref{neural_network}, left-hand side).  In the second case, the output provides
the coefficients of the series (\ref{series}) instead      (see Fig. \ref{neural_network},
right-hand side). In this work,       both representations have been utilized and we have
observed that,         even if they both practically provide the same results, the second
representation is obviously faster to converge (as expected). Moreover, in both cases the
hidden layer can have as many units as the computational resources allow.

As a final comment, the Reader should also note that these approaches are suitable    for
both one-, two- and three-dimensional spaces since the structure of the network is simple
and does not require important resources. As a matter of fact, these approaches have been
tested in these dimensionalities and have always provided the right answers      (see the
numerical experiments discussed below).


\section{Energy Minimization Principle}

It is relatively easy to show that,                by algebraically manipulating equation
(\ref{schroedinger_time_independent}),        the total energy of a quantum system can be
analitically expressed as:
\begin{equation}
 E = \frac{\int_0^L dx \psi^*(x) \hat{H} \psi(x)}{\int_0^L dx \psi^*(x) \psi(x)}
\label{total_energy}
\end{equation}
In the context of    representing wave functions by means of  feedforward neural networks,
one can quickly see that        the total energy becomes a function of the weights of the
network, i.e.
\begin{equation}
 E=E({\bf{w}}),
\end{equation}
where by ${\bf{w}}$ one intends the set of weights and biases of the network.    Moreover,
in order to enforce the wave function to have a value equal to zero at the boundaries, in
other words closed boundary conditions,              an extra term is added to the energy
(\ref{total_energy}) which reads:
\begin{equation}
 E=E({\bf{w}}) + \lambda \times [ \psi^*({\bf{w}}, 0) \psi({\bf{w}}, 0) + \psi^*({\bf{w}}, L) \psi({\bf{w}}, L) ]
\end{equation}
(with $\lambda$ a constant) which naturally   ensures that the wave function is decreased
at the boundaries of the spatial domain. An fact of paramount importance to note is  that
the integral involved in      the total energy (\ref{total_energy})     can be calculated
analitically when approached by a single-hidden-layer       neural network with tractable
activation functions.      This fact seems to suggest that we might be able to completely
by-pass the problem of the sign which affects very advanced methods   such as the quantum
Monte Carlo one \cite{Carleo-2017}.

Since we are interested in finding the ground state of a system, our strategy consists in
simply varying the weights of the network representing the wave function until          a
(hopefully global) minimum is reached.        This can be achieved in different ways, for
instance by means of Monte Carlo importance sampling   and gradient descent. In this work,
we utilize the covariance matrix adaptation evolution strategy (CMA-ES),         which in
contrast with the vast majority of evolutionary algorithms,       is quasi parameter-free,
while keeping the very useful feature of being embarassingly parallelizable.     Moreover,
the CMA-ES has been empirically successful in hundreds of applications \cite{CMA}.    For
the sake of completeness, the main tenets of this method are provided below    and the
interested Reader can refer to \cite{CMA} for more details.

\bigskip

{\sl{CMA-ES genetic strategy}}. Evolution strategies are stochastic,      derivative free
methods with applications in non-linear or non-convex continuous    optimization problems.
Such methods belong to the class of optimizers known as evolutionary algorithms which are
broadly based on the principle of biological evolution. In more specific details, at each
iteration new individuals are randomly generated (i.e. candidate solutions denoted by $x$)
from current parental individuals. Then, the best individuals are selected to become,  in
turn, the parents for the next iteration based on their fitness or objective     function
value $f=f(x)$ (which, in our specific case, consist of the total energy of the    system
(\ref{total_energy})). In this way, a sequence of individuals is generated,           and
individuals with smaller and smaller energy are generated.  Such approach is particularly
useful in situations where the function $f=f(x)$ is ill-conditioned (which,  for instance,
cannot be treated by deterministic algorithms).

Two  main  precepts    for  the  adaptation  of  parameters  are  exploited in the CMA-ES
algorithm. The first one consists of a maximum-likelihood principle which is based on the
idea of increasing the probability of successful candidate solutions. A covariance matrix
of the distribution is accordingly updated        which guarantees that the likelihood of
previously successful search steps is increased. The second one consists of recording two
paths of the time evolution of the distribution mean of the strategy,    called search or
evolution paths. The evolution paths are exploited in two ways. One path is used  for the
covariance matrix adaptation procedure.   The other path is used to conduct an additional
step-size control. In particular,    the step-size control effectively prevents premature
convergence while allowing fast convergence to an optimum solution.

\begin{figure}
\centering
\includegraphics[width=0.75\textwidth]{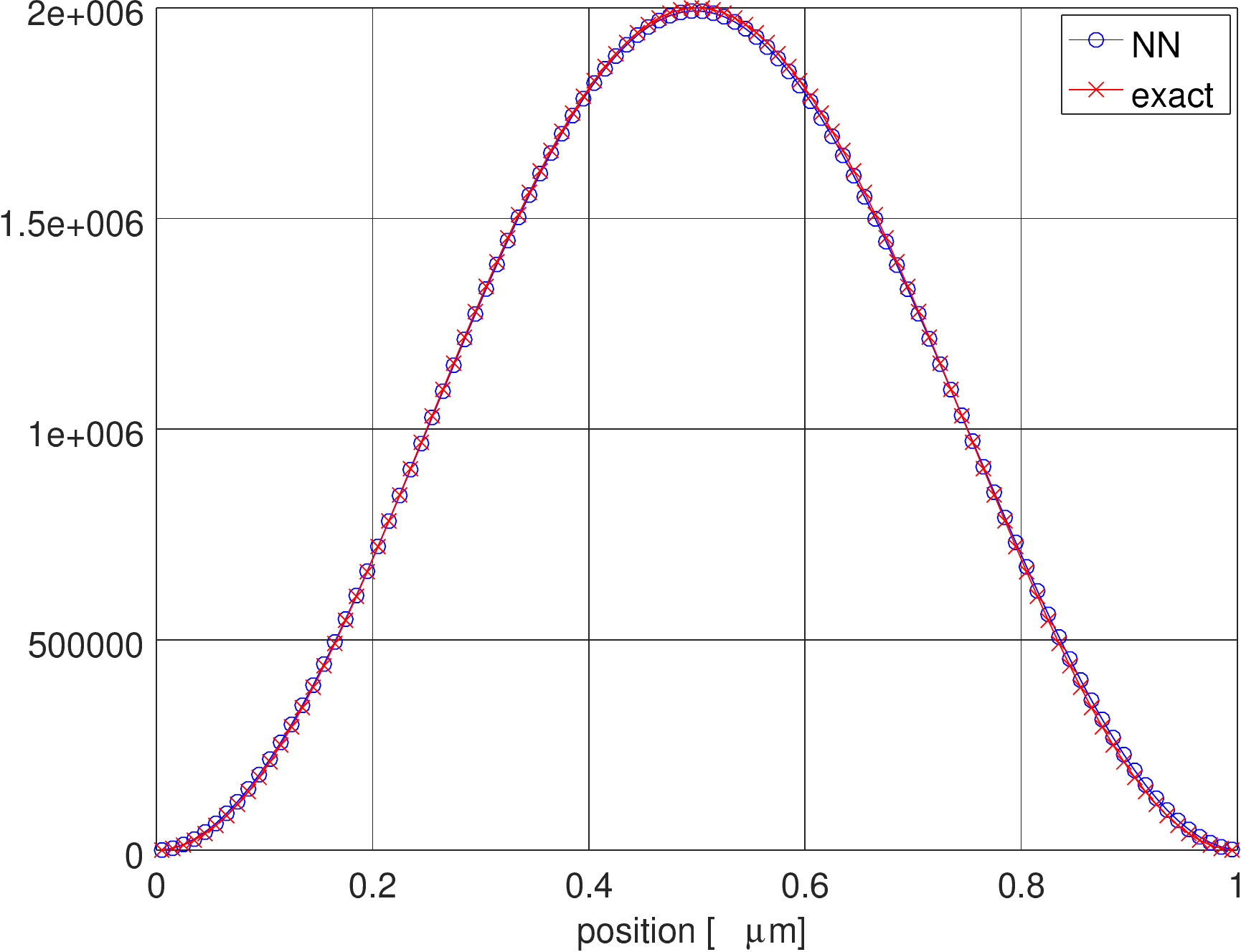}
\caption{Comparison between the exact solution - (red) cross curve - and the neural network based approach - (blue) circle curve - for the one-dimensional
particle in a box problem. Although this is a seminal work, good agreement can already be observed between the numerical and exact
probability densities.}
\label{particle_in_a_box_1D}
\end{figure}

\section{Numerical Validation}

We now present a series of validation tests to show that the method suggested     in this
paper, although in a preliminary shape,      is already in good agreement with the theory.
Specifically, these tests consist in finding the ground state of systems made of a single
or two-electrons, and compare it with exact or numerically available solutions.       The
systems taken into account are: one or more electrons {\sl{1)}} in a closed box, {\sl{2)}}
inside a finite potential well,       and {\sl{3)}} in the presence of a single potential
barrier. The numerical details and results are presented in the rest of this section.

An interesting (heuristic) point to keep in mind is that, in all experiments presented in
this section, the CMA-ES genetic algorithm has never ended in a local extremum.    It has
always converged to the ground state of any system that has been approached       in this
seminal work. Although more investigation is still needed,       it represents an initial
encouraging sign that it might be utilized in more complex situations.      Moreover, all
experiments described in this section have been approached     by networks with only $16$
hidden units and, yet, they were able to provide the correct answer. To the author,  this
is a clear indication that this approach might actually be able to reduce  the complexity
of the Hilbert space.

\bigskip

{\sl{Particle in a box}}.         The particle in a box model, also known as the infinite
potential well, describes a  free  particle in a finite domain surrounded by two infinite
barriers.        This problem is one of the very few in quantum mechanics which exact (or
analytical) solution is known.  Due to its simplicity,   this situation represents a good
initial benchmark test to validate our approach. The potential energy in this model reads
(for simplicity we hereby refer to the one-dimensional case, its generalization to higher
dimensions being trivial):
$$
 V(x) = 0, \forall x \in D,
$$
where $D$ is a finite spatial domain represented by an interval $[0, L]$.  It is possible
to show that,     in this case,                    the exact solution of the eigenproblem
(\ref{schroedinger_time_independent}) reads:
\begin{equation}
 \psi_n(x) = \sqrt{\frac{2}{L}} \sin{(k_n x)},
\end{equation}
with $k_n = \frac{n \pi}{L}$, and
\begin{equation}
 E_n = \frac{n^2 \pi^2 \hbar^2}{2 m L^2},
\label{particle_in_a_box_energy}
\end{equation}
where $m = 9.10938356 \times 10^{-31}$ Kg is the mass of the particle        (i.e. a free
electron). The ground state is, obviously, represented by the case $n = 1$.

Below, we report the results of applying the approach discussed in this work along with a
comparison with the exact solution.             A good agreement can be observed (see Fig.
\ref{particle_in_a_box_1D}) for a domain with $L = 1\mu$m.  The exact ground state energy
can be    computed by means of formula (\ref{particle_in_a_box_energy})   and,    in this
particular case, is equal to $3.7603 \times 10^{-7}$ eV.   The energy found by our method
is equal to $3.7254 \times 10^{-7}$ eV, in good agreement with the theory.

\begin{figure}
\centering
\includegraphics[width=0.75\textwidth]{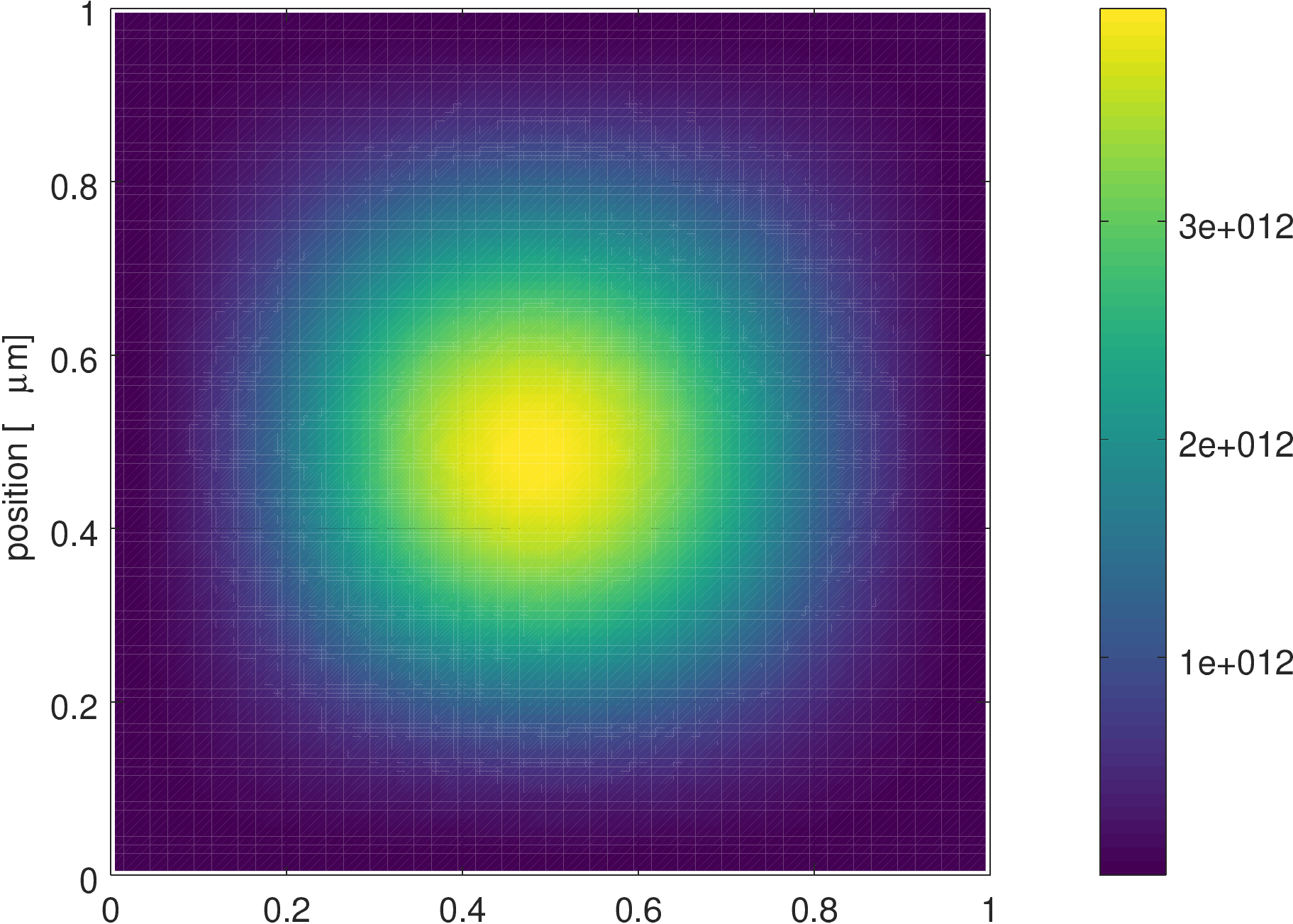}
\caption{Ground state of a two-dimensional particle in a box found by means of the neural network based approach.
One can clearly observe how the expected symmetries of the ground state are respected by this method.}
\label{particle_in_a_box_2D}
\end{figure}

\begin{figure}
\centering
\includegraphics[width=0.3\textwidth]{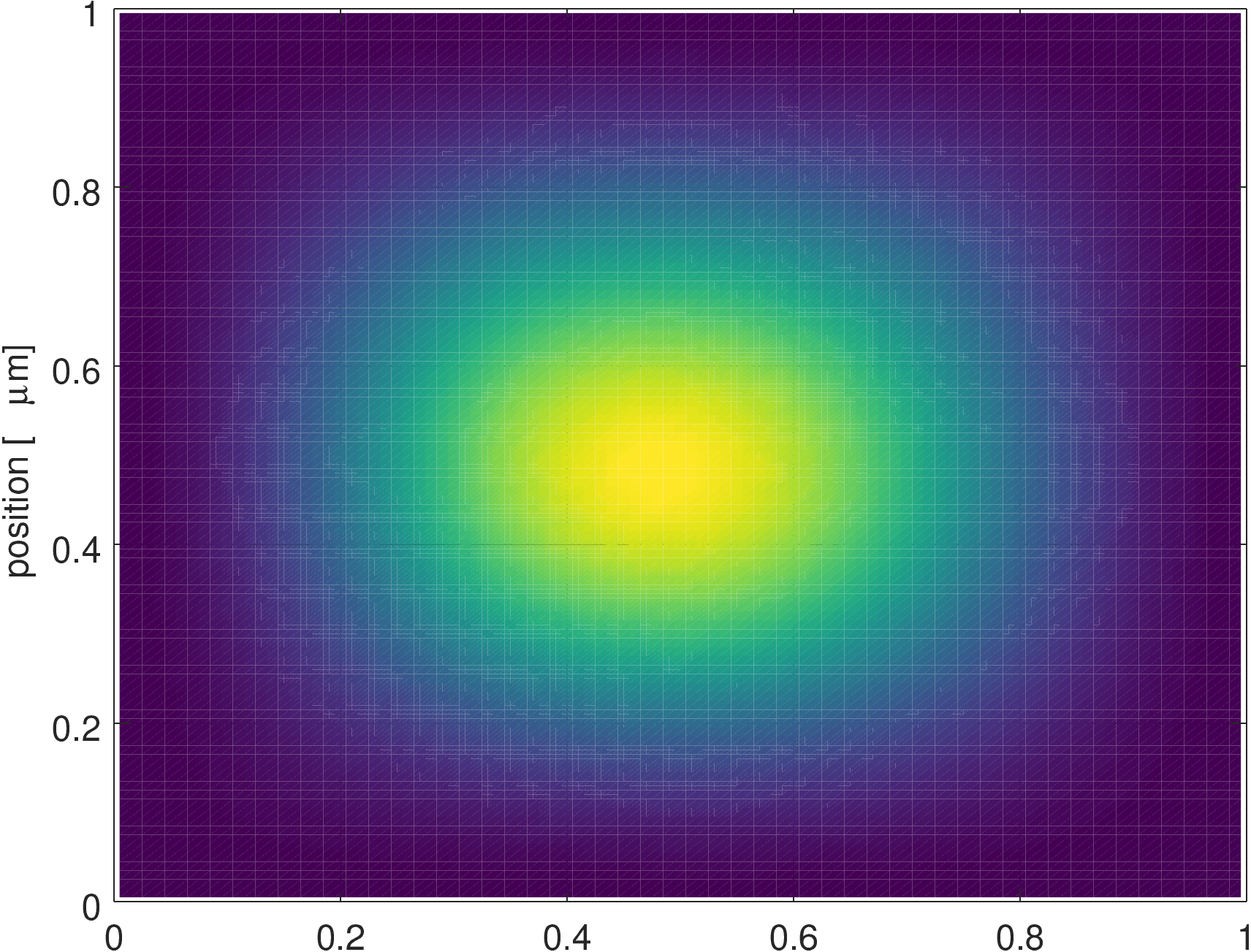}
\includegraphics[width=0.3\textwidth]{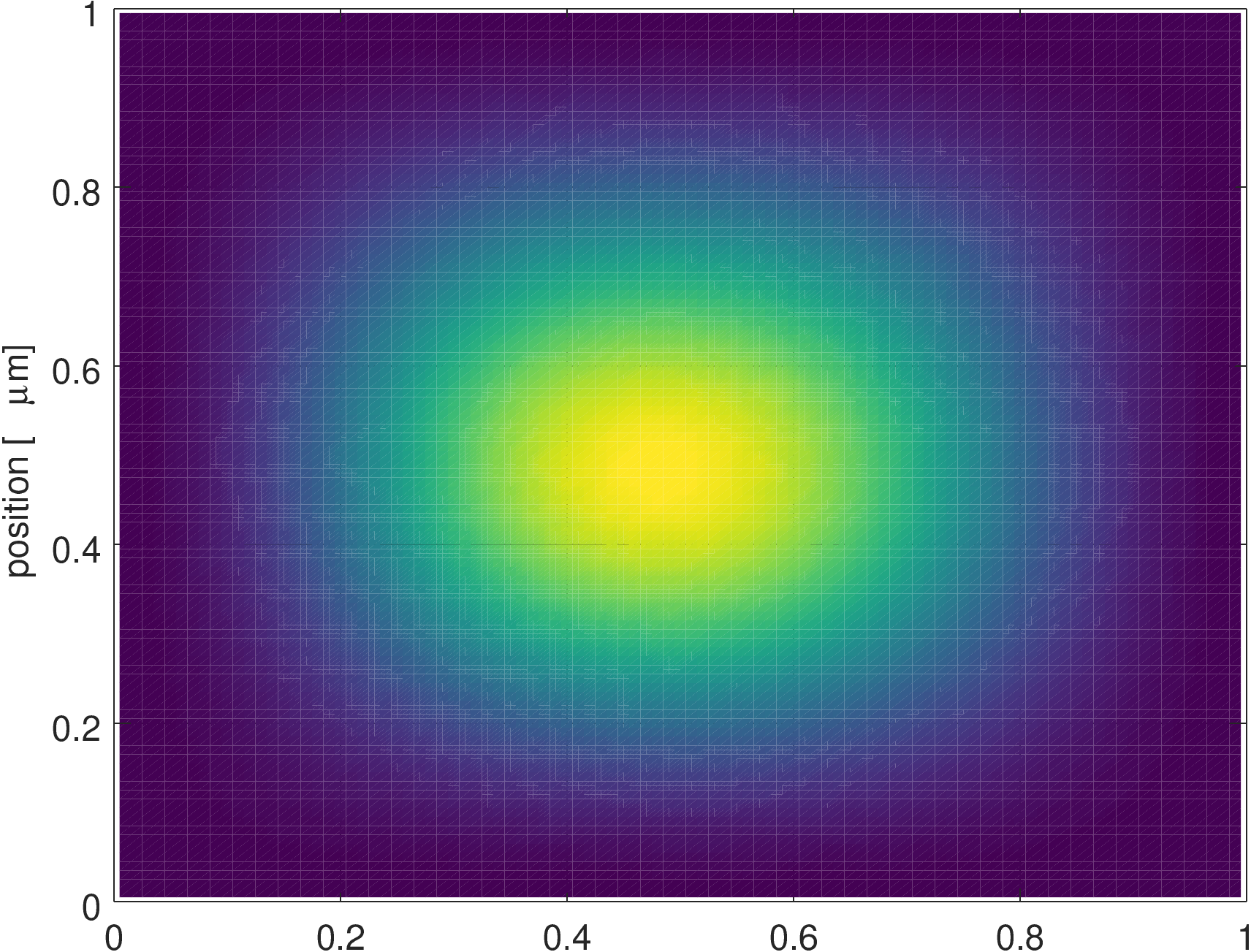}
\includegraphics[width=0.3\textwidth]{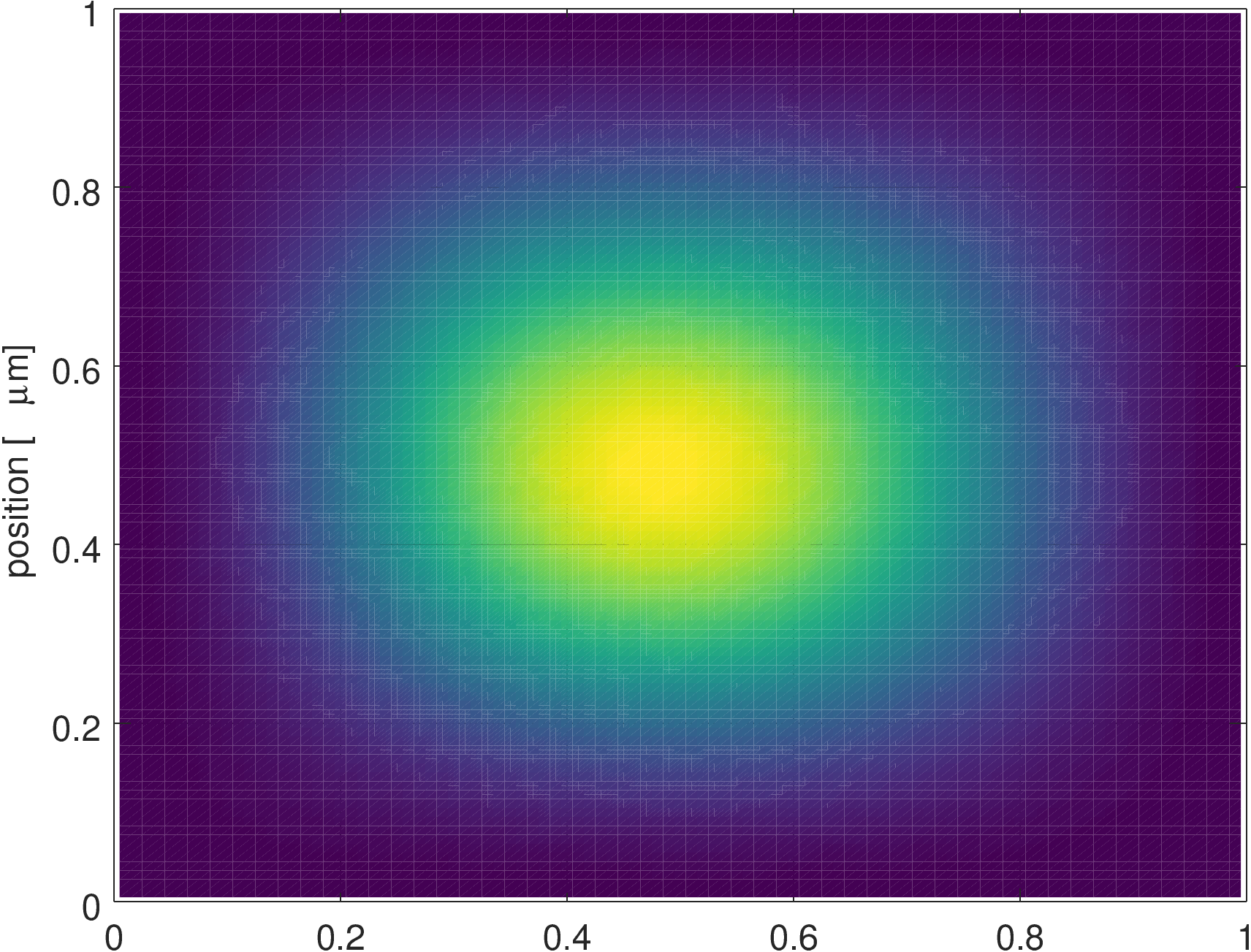}
\caption{Cut of the three-dimensional ground state for a particle in a box computed by means of a feedforward neural network.
The cuts are performed on the planes $x=\frac{L_x}{2}$ (left), $y=\frac{L_y}{2}$ (middle) and $z=\frac{L_z}{2}$ (right) respectively.
The symmetry is very well respected on every plane.}
\label{particle_in_a_box_3D}
\end{figure}

A similar test is then   performed in a two dimensional space as well    (with dimensions
$1 \mu$m $\times 1 \mu$m) and Fig. \ref{particle_in_a_box_2D} shows the ground state wave
function obtained. One can clearly observe that the expected symmetry of the ground state
is respected. The energy found is equal to $7.40031 \times 10^{-7}$ eV   which is in good
agreement with the theoretical value of $7.52061 \times 10^{-7}$ eV.    Finally, the same
test is performed in a three dimensional space with dimensions   $1 \mu$m $\times 1 \mu$m
$\times 1 \mu$m.        The ground state energy is, in this case, expected to be equal to
$1.12809 \times 10^{-6}$ eV  while we obtain a value equal to $1.11548 \times 10^{-6}$ eV,
again, in good agreement with the theory. Fig. \ref{particle_in_a_box_3D} shows the  cuts
of the three-dimensional ground state on the planes $x=\frac{L_x}{2}$,  $y=\frac{L_y}{2}$
and $z=\frac{L_z}{2}$. A very good symmetry structure can be observed.

\begin{figure}
\centering
\includegraphics[width=0.75\textwidth]{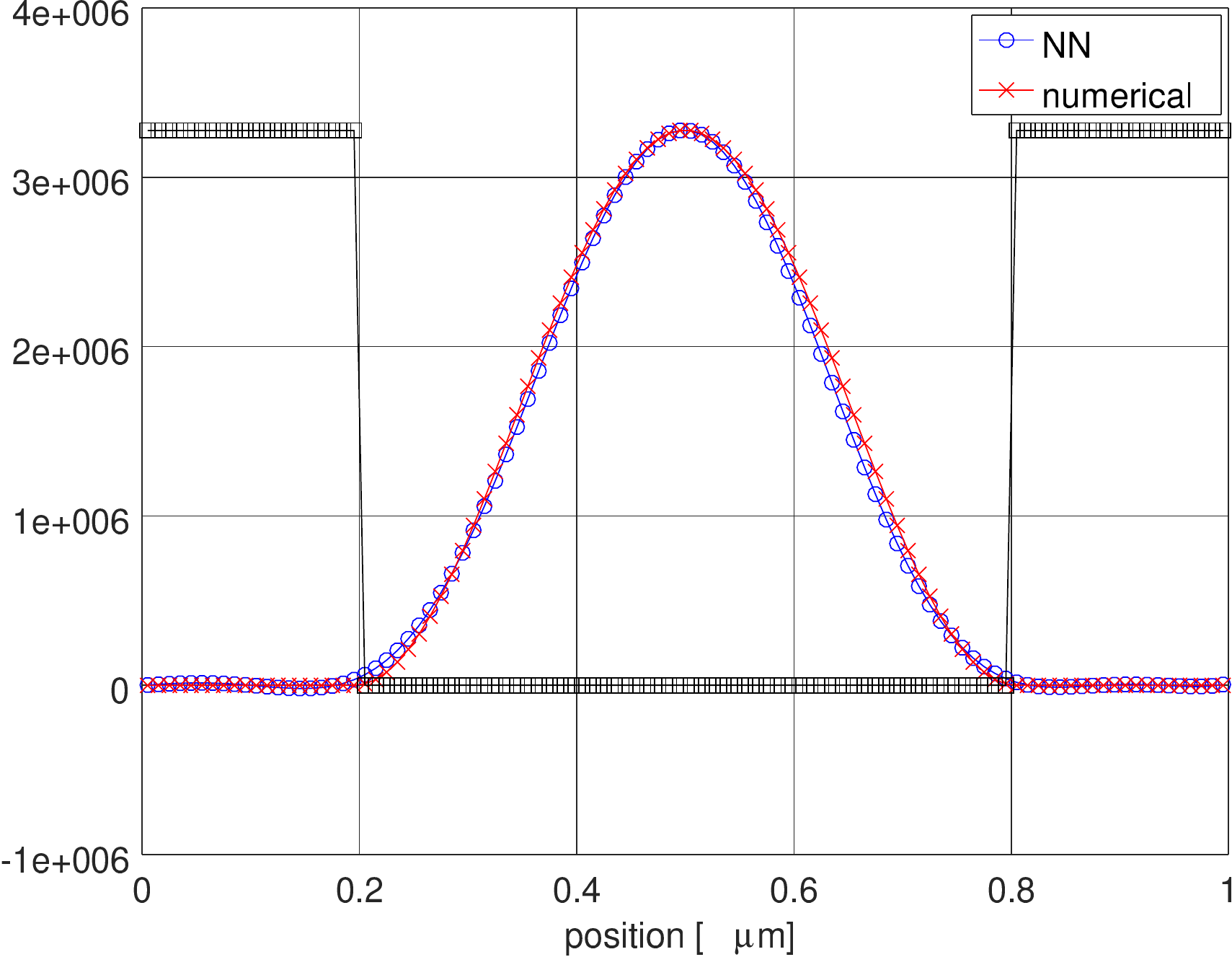}
\caption{Comparison between a numerically computed solution - (red) cross curve - and the suggested neural
network based approach - (blue) circle curve - for the one-dimensional
particle in a finite potential well - (black) square curve. A good agreement can be observed.}
\label{finite_potential_well_1D}
\end{figure}

\bigskip

{\sl{Finite potential well}}.   The finite potential well is a typical problem of quantum
mechanics which can be seen  as a generalization of the particle in a box problem.     It
consists of two barriers placed  at     the right-hand and left-hand sides of the spatial
domain. In such context,   if the total energy of the particle is smaller than the energy
of the barriers then it cannot be found outside of the box.       The results obtained by
exploiting a neural network representation of         the quantum state are compared to a
deterministic eigensolver \cite{eigen} implemented    in the LAPACK library \cite{lapack},
and reported in     Fig. \ref{finite_potential_well_1D}. The computed numerical value for
the energy is equal $3.2363 \times 10^{-6}$ eV       while the value found by the network
approach is equal to $3.2872 \times 10^{-6}$ eV.     It is clear that a good agreement is
reached between two very different approaches.

\bigskip

\begin{figure}
\centering
\includegraphics[width=0.75\textwidth]{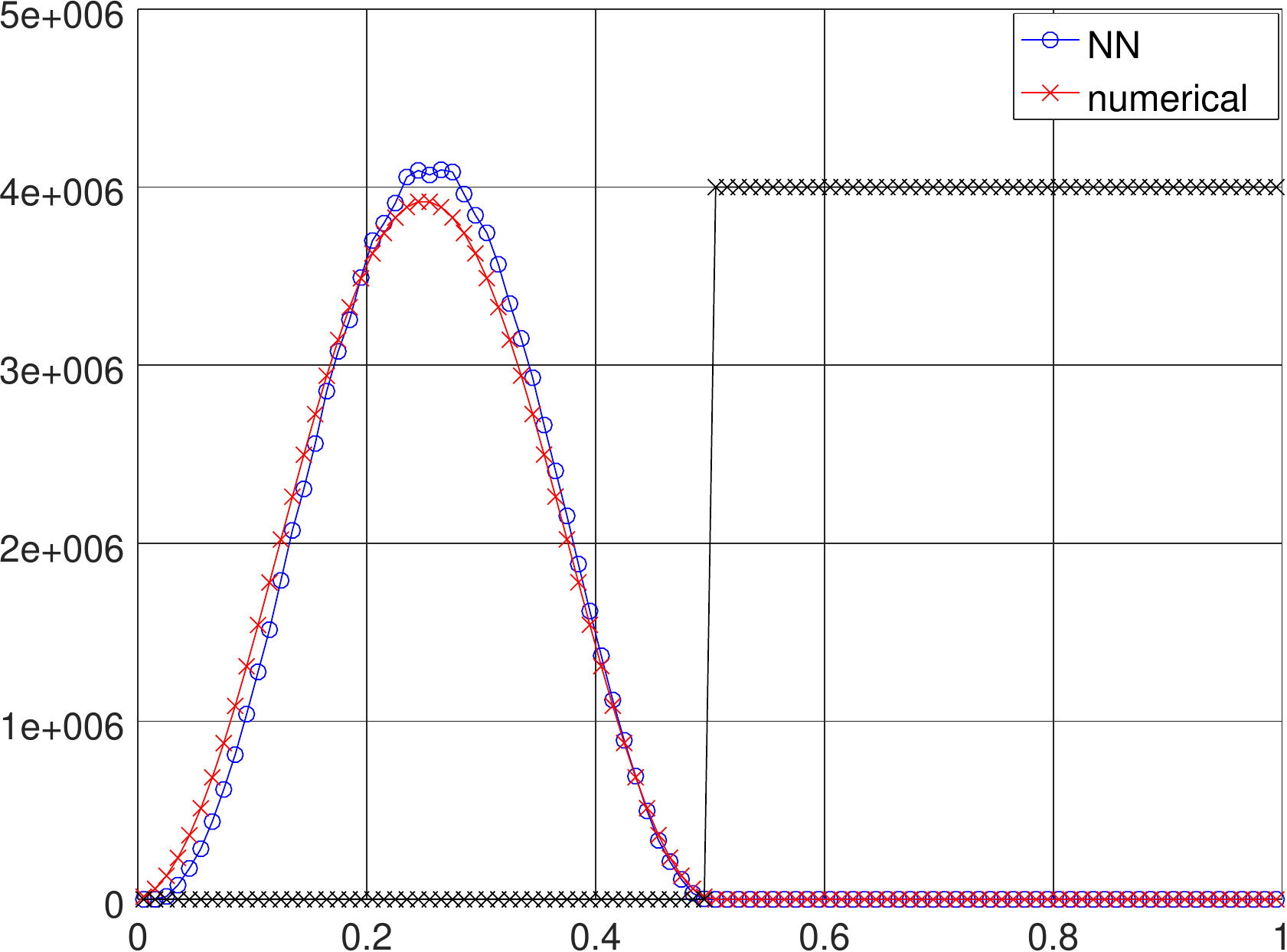}
\caption{Comparison between the exact solution - (red) cross curve - and the neural network based approach - (blue) circle curve - for the one-dimensional
single potential barrier problem. Some differences are noticeable between the numerical and exact probability densities.}
\label{single_potential_barrier_1D}
\end{figure}

{\sl{Single step barrier}}. In quantum mechanics,  the single step potential barrier is a
standard problem  which illustrates the phenomena of quantum tunneling and reflection. In
practice, it consists of solving the time-independent Schr\"{o}dinger            equation
(\ref{schroedinger_time_independent}) in the presence of a potential which reads:
$$
 V(x) = V_0, \forall x \in [a, b],
$$
and $0$ elsewhere for some value $V_0 \neq 0$ and $0 \le a < b \le L$. In particular, for
the results presented here, we used $a=\frac{L}{2}$, $b=L$ and $V_0 = -0.1$ eV.  The same
domain length as the previous test is utilized.    The comparison between the numerically
computed solution and the one based                    on our approach can be seen in Fig.
\ref{single_potential_barrier_1D}. The agreement remains satisfying.   In fact, the peaks
are found in the same position,   and the shape of the two wave functions  is practically
the same, although some differences between the two solutions can be observed. This could
be due to the very different nature of the two methods,  being one essentially stochastic
and the other deterministic (further investigations are required). In any case,      this
initial agreement remains promising.         Energy-wise, the value found by the standard
eigensolver is equal to $1.4450 \times 10^{-6}$ eV while the    one found with our neural
network approach  is equal to $1.5872 \times 10^{-6}$ eV, another indication that the two
methods are in acceptable agreement.

\bigskip

{\sl{Non-interacting particles}}.      An important validation test is represented by the
simulation  of a quantum system made of $N$ non-interacting electrons.  In fact, although
it involves a very particular situation, it is one of the few many-body systems     which
exact solution is known and which can be simply written as the product of single-electron
wave functions (Hartree products).         In more mathematical details, such a system is
characterized by an Hamiltonian which reads:
\begin{equation}
 \hat{H}(x_1, \dots, x_n) = \sum_{i=1}^{n} \hat{h}_i = \sum_{i=1}^n [ -\frac{\hbar^2 \partial}{2 m_i \partial x_i} - q V(x_i)],
\label{Hartree_Hamiltonian}
\end{equation}
where $\hat{h}_i$ is an operator acting on the $i$-th particle only,    $m_i$ is its mass
and $q$, as usual, its charge.  The exact solution for this problem is well known and can
be written in the shape of an Hartree product.         The Hartree product is a many-body
wave function, given as a combination of wave functions of the individual particles.   It
assumes that the particles are independent and,    therefore,            is unsymmetrized.
Mathematically, a two-body Hartree product reads:
$$
 \psi(x_1, x_2) = \psi_1(x_1) \psi_2(x_2),
$$
where $\psi_1=\psi_1(x_1)$ and $\psi_2=\psi_2(x_2)$ are the solutions of the problem
$$
 \hat{h}_i \psi_i(x) = \epsilon_i \psi_i(x).
$$
Finally,        it can be proven that the total energy of the problem is equal to the sum
$\epsilon_1 + \epsilon_2$. This provides a simple way to check the validity of the energy
found by the method presented in this work.             Moreover, the probability density
$n = n(x_1, x_2)$ in the configuration space reads:
\begin{eqnarray}
n(x_1, x_2) &=& \psi^*(x_1, x_2) \psi(x_1, x_2) \nonumber \\
&=&  \psi^*_1(x_1) \psi^*_2(x_2) \psi_1(x_1) \psi_2(x_2) \nonumber \\
&=& \psi^*_1(x_1) \psi_1(x_1) \psi^*_2(x_2) \psi_2(x_2) \nonumber \\
&=& n_1(x_1) n_2(x_2) \nonumber
\end{eqnarray}
which clearly highlights the fact that the two-body density $n=n(x_1,x_2)$    is equal to
the product of the two one-body densities   $n_1=n_1(x_1)$   and $n_2=n_2(x_2)$.     This
provides a simple way to check the validity of the wave functions found by our approach.

\begin{figure}
\centering
\includegraphics[width=0.75\textwidth]{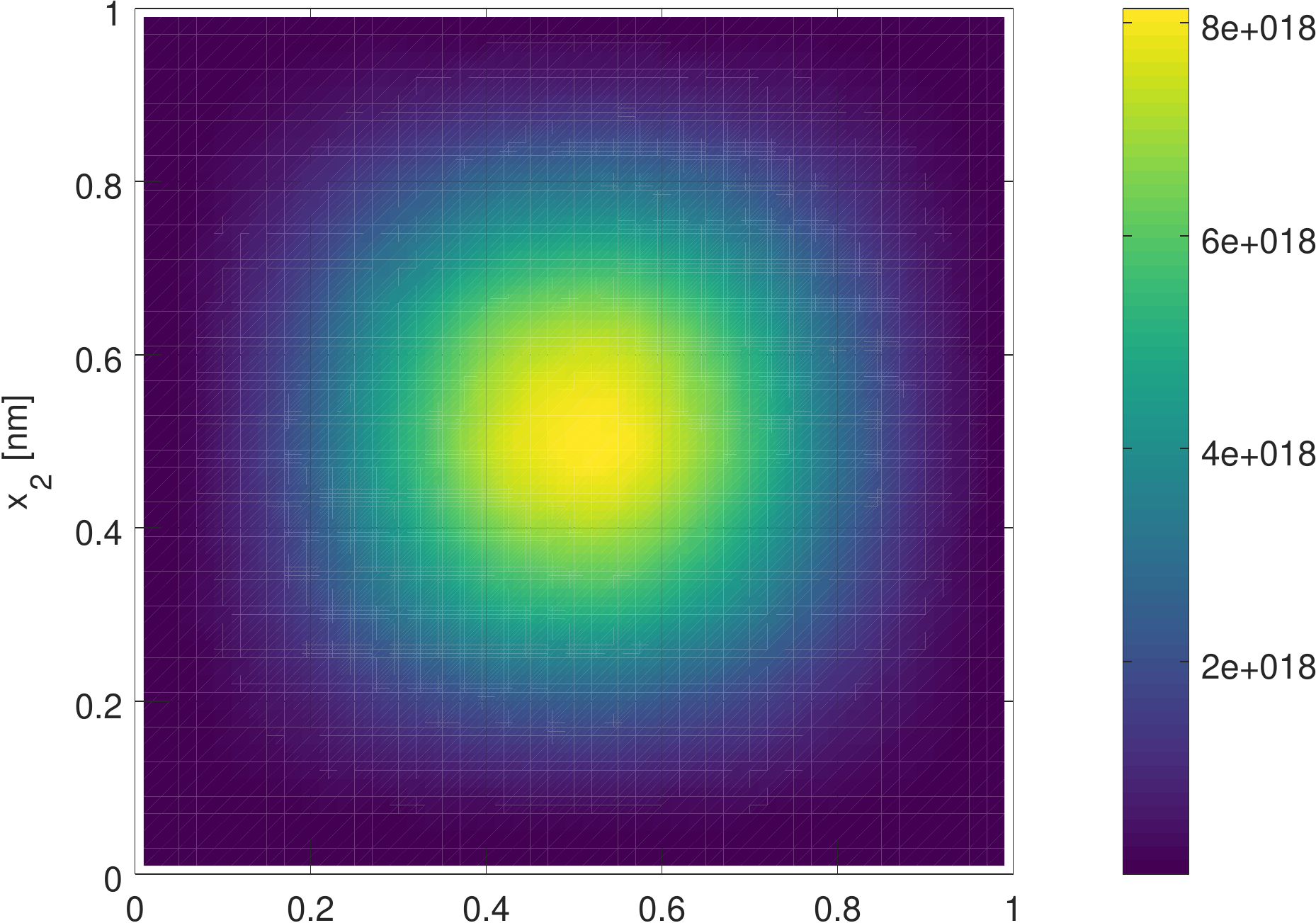}
\caption{Solution of the 2-body particle-in-a-box problem in the two-dimensional configuration space. A very good symmetry can be observed,
in good agreement with the theoretical solution.}
\label{particle_in_a_box_2_body}
\end{figure}

\begin{figure}
\centering
\includegraphics[width=0.45\textwidth]{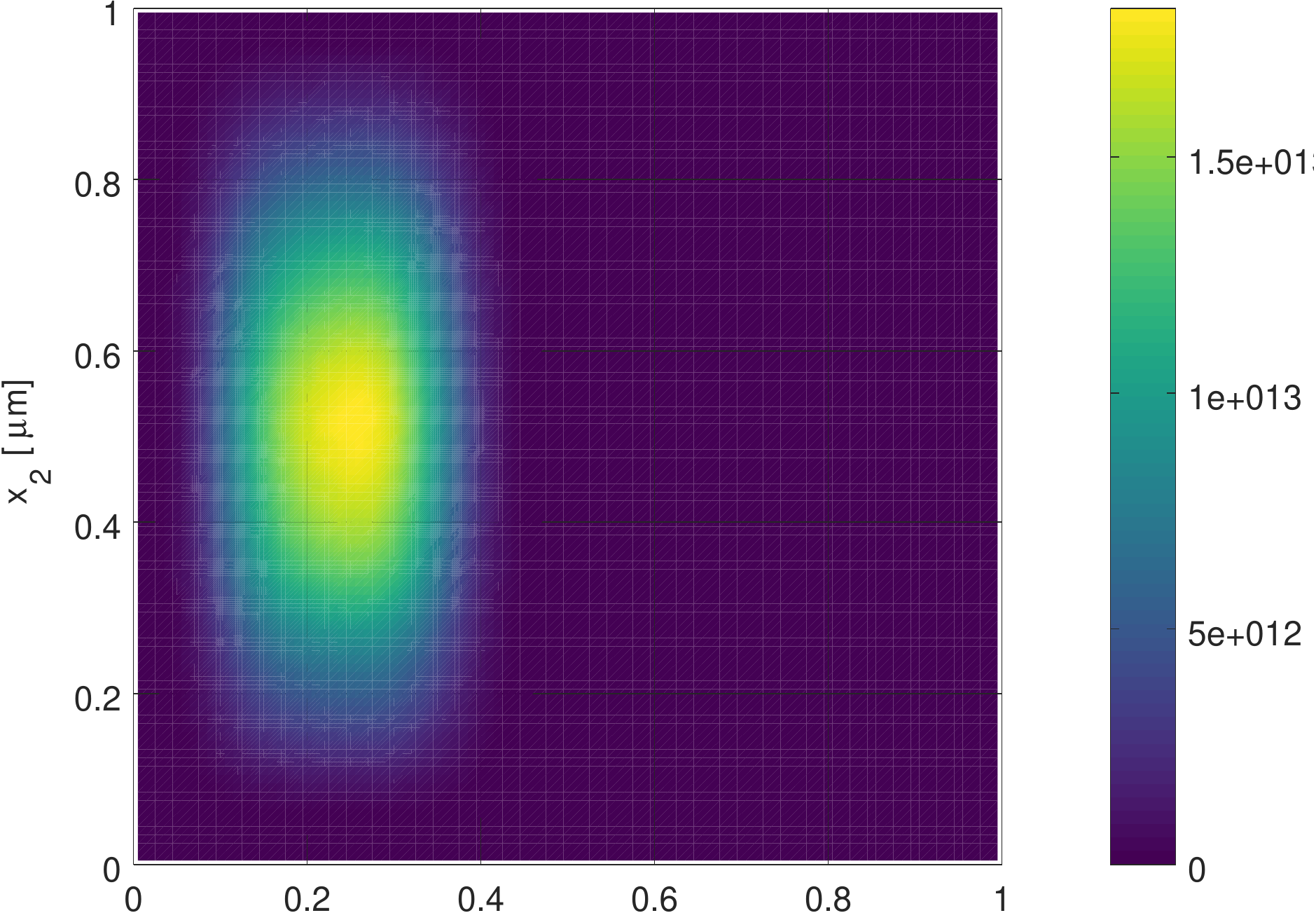}
\includegraphics[width=0.45\textwidth]{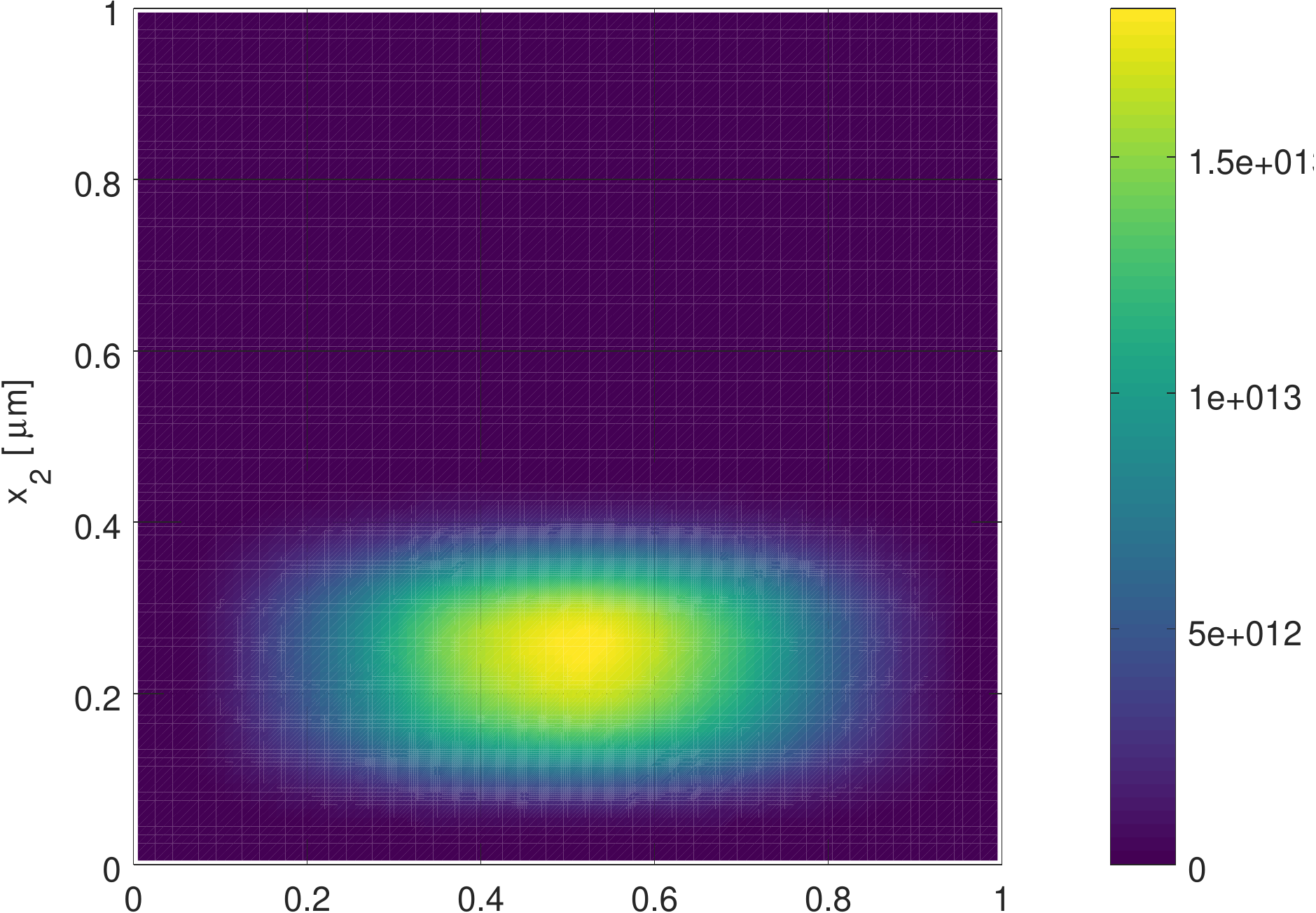}
\caption{Two solutions of the 2-body problem in the two-dimensional configuration space corresponding to a particle which feels a step barrier and
one which does not feel any potential. A very good symmetry can be observed, in good agreement with the theoretical solution.}
\label{Hartree_2}
\end{figure}

\begin{figure}
\centering
\includegraphics[width=0.75\textwidth]{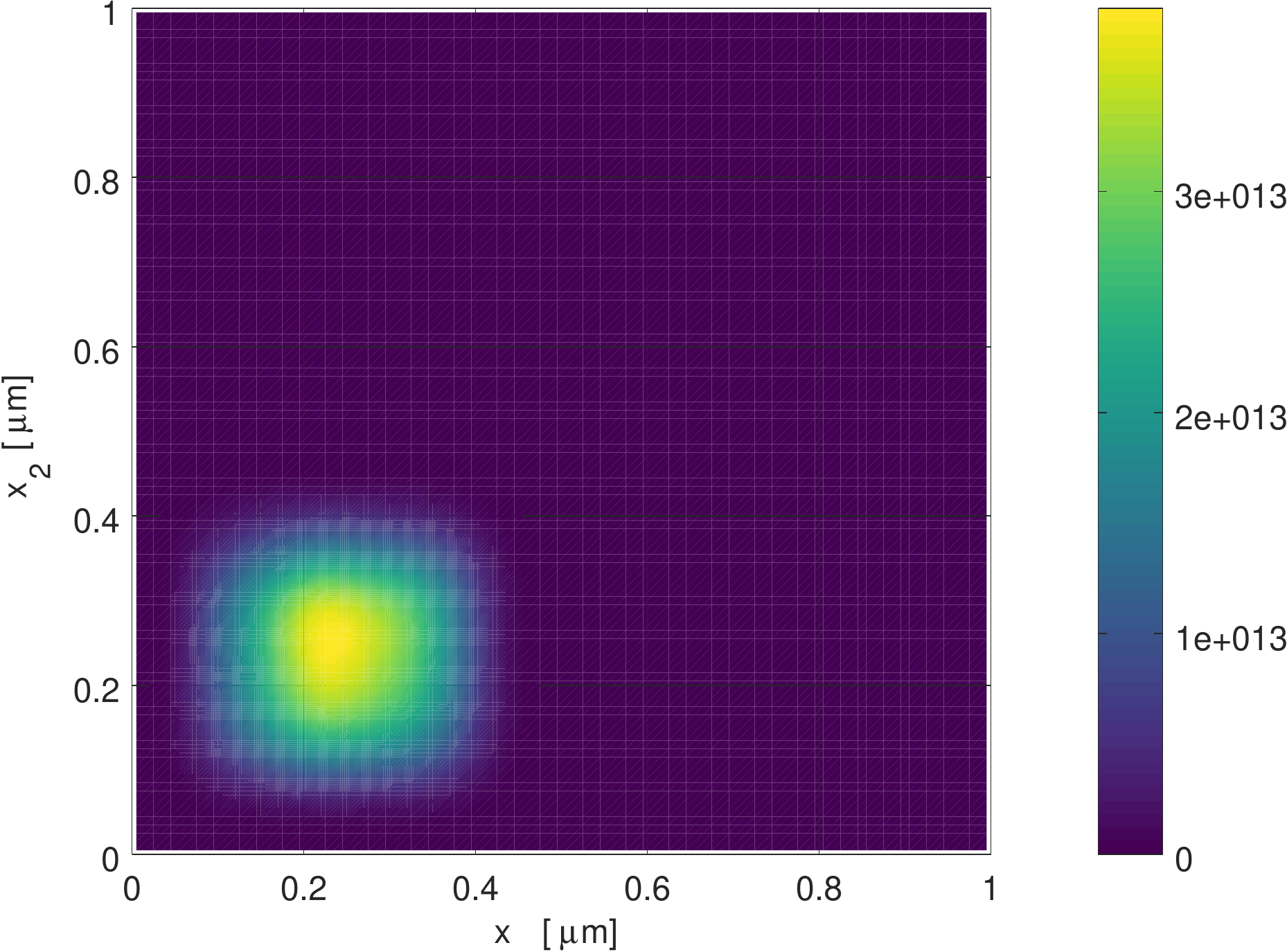}
\caption{Two solutions of the 2-body problem in the two-dimensional configuration space corresponding to two non-interacting particles
in the presence of a step barrier.}
\label{Hartree_3}
\end{figure}

In this validation test, the spatial domain is now equal    to      $[0, 1.0]$ nm and the
theoretical expected energy is equal to the sum of the energies of the single   particles,
i.e. $0.7520$ eV. The value found with our approach is equal to $0.7537$ eV  in excellent
agreement with the theory. Fig. \ref{particle_in_a_box_2_body}.       We then performed a
second numerical experiment still involving two non-interacting particles, one feeling no
potential and the other feeling a step potential barrier in the domain    $[0, 1.0] \mu$m.
In this situation, the energy is expected to be lower than $3.7603 \times 10^{-7} +  1.5041 \times 10^{-6} = 1.8801 \times 10^{-6}$ eV
(which corresponds   to the energy of a particle confined to the domain length $1.0 \mu$m
and $0.5 \mu$m respectively). Our approach finds a value equal to $1.8603 \times 10^{-6}$
eV. Fig. \ref{Hartree_2} shows the squared magnitude of the computed        two-body wave
functions defined over  the configuration space where one electron feels a step potential
barrier and the are does not, and vice-versa. Clearly,          the shape of the function
corresponds to the one expected.        A final test is presented in Fig. \ref{Hartree_3}
showing the probability density for two electrons both feeling the presence of a     step
potential barrier. The shape of the wave function is, again, the one expected. The energy
expected in such system is equal to $2 \times 1.5041 \times 10^{-6} = 3.0082 \times 10^{-6}$ eV
while the value found with by our approach     is equal to $3.2741 \times 10^{-6}$ eV, in
good agreement with what expected.

\bigskip

{\sl{Interacting particles}}.     This final validation test represents an important step
since it involves exchange effects happening in quantum systems of Fermions.      In this
context,                        the Hamiltonian of such systems cannot be expressed as in
(\ref{Hartree_Hamiltonian})     since they contain an extra term modeling the interaction
between particles and which, for the case of electrons, reads:
$$
 V(x_1, x_2) = \frac{q^2}{4 \pi \epsilon_0} \frac{1}{|x_1 - x_2|},
$$
and where the constant $\epsilon_0$ is the permittivity of vacuum.   One should note that
in every numerical experiments performed in the context of interacting particles,     one
expects the total   energy   of the system to be always bigger     than the one involving
non-interacting particles,        due to the fact that interacting particles are strongly
confined not only by the dimensions of the box but also by        the Coulombic potential
created by the other particles in such system.  This is clearly observed from the results
obtained by the approach suggested in this work.

\begin{figure}
\centering
\includegraphics[width=0.45\textwidth]{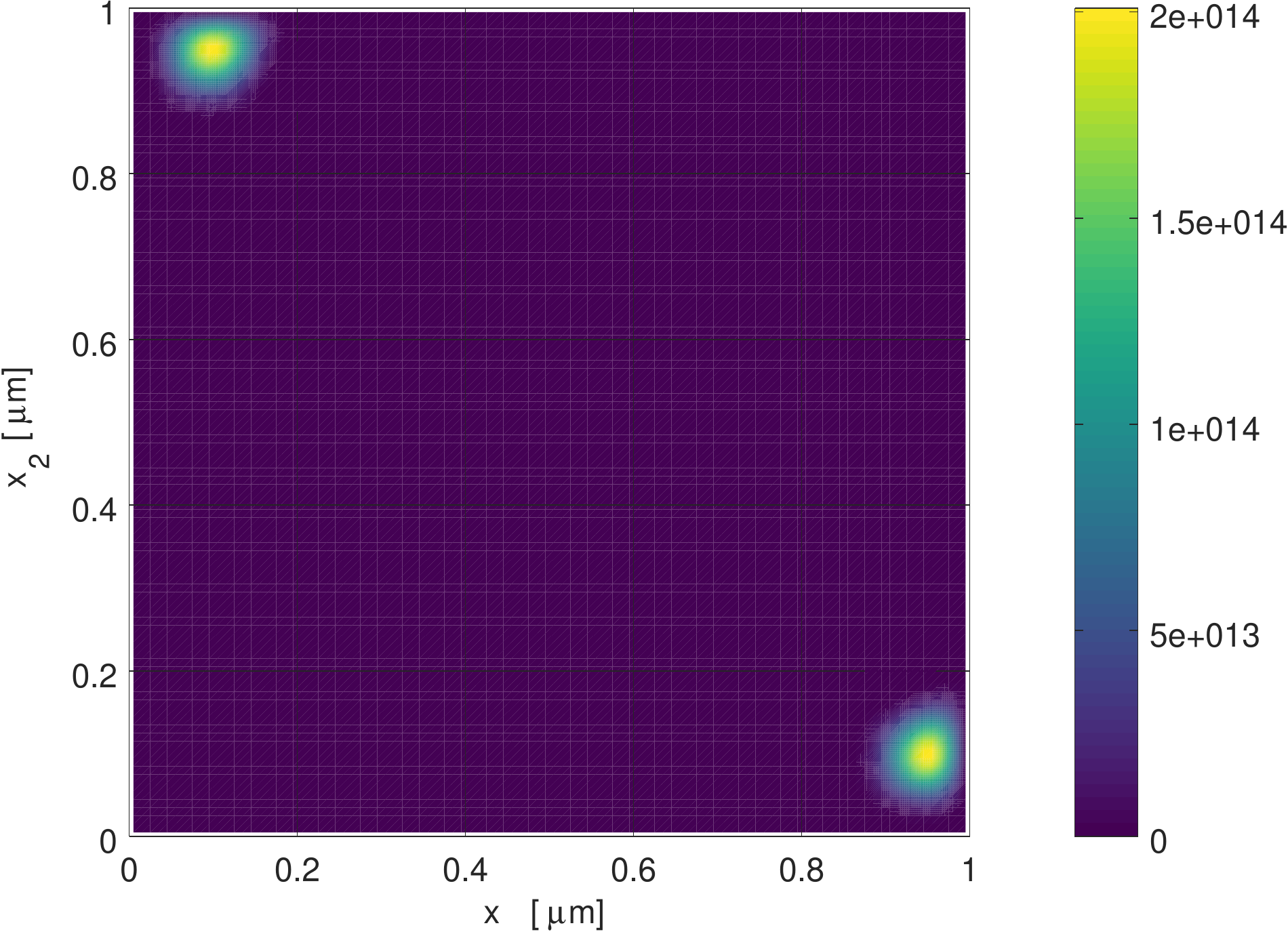}
\includegraphics[width=0.45\textwidth]{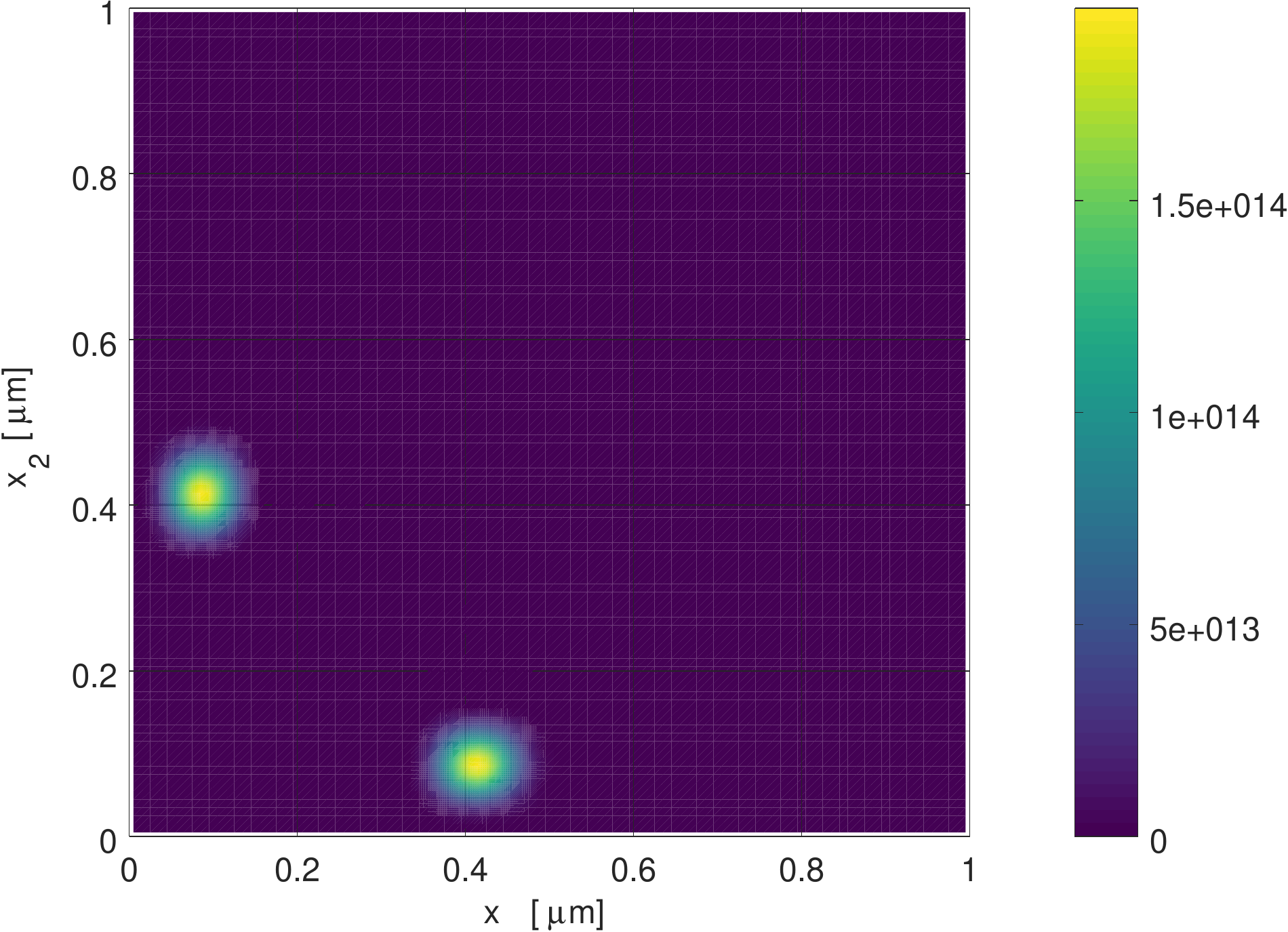}
\caption{Probability density functions of two interacting Fermions in a empty box (left-hand side)
and in the presence of a step potential barrier (right-hand side).}
\label{interacting_fermions}
\end{figure}

The tests consist of finding the ground state of two interacting particles    in an empty
box first (i.e. two interacting particles in a box), and, then, in the presence of a step
potential barrier.    The energies found in both cases are $3$ orders of magnitude bigger
than their corresponding non-interacting counterparts    (i.e. $1.7619 \times 10^{-3}$ eV
and $4.6467 \times 10^{-3}$ eV respectively). The probability density functions for these
tests are shown in Fig. \ref{interacting_fermions}  (left-hand side is the empty box case,
right-hand side corresponds to particles in the presence of a step potential     barrier)
where the confinement of the electrons is clearly visible (as expected).

\section{Conclusions and Future Works}

In this paper, a new method to represent the state of a quantum system,      based on the
universal representation theorem has been suggested. The method is similar, in its tenets,
to the one discussed in \cite{Carleo-2017} but applies to systems beyond spinglasses  and
exploits feedforward neural networks rather than Boltzmann machines.   Several validation
tests, involving interacting and non-interacting electrons have been presented which hold
the promise of more complex simulations.    Many directions still need to be investigated
though. For instance, nothing has been said about  the non-linear activation functions of 
the hidden layer   and       it would be of high interest to see what the effects of such
non-linearities (ReLU, etc.) would be on the accuracy of the ground state.       The same
question could be asked for the depth and width of the network (in this work we have been
using single-hidden layer neural networks only).   In the same way,       the plethora of
sampling methods available should be explored to clarify if any further advantage can  be
obtained. These will be the subject of next future works.

%

\bigskip

{\bf{Acknowledgments}}. The author would like to thank G.~Marceau~Caron and
S.~Blackburn for their very fruitful and valuable conversations. A special thanks goes to M.~Anti for her loving support and encouragement.


\begin{thebibliography}{1}

\bibitem{Carleo-2017}
G.~Carleo, M.~Troyer,
Solving the Quantum Many-body Problem with Artificial Neural Networks,
Science 355, pp. 602–606, (2017).

\bibitem{Carleo-2018}
G.~Carleo, Y.~Nomura, M.~Imada,
Constructing Exact Representations of Quantum Many-body Systems with Deep Neural Networks,
Nature Communications 9, 5322, (2018).

\bibitem{Kolmogorov}
A.N.~Kolmogorov,
On the Representation of Continuous Functions of Several Variables by Superposition of Continuous Functions of one Variable and Addition,
Doklady Akademii. Nauk USSR, 114, pp. 679-681, (1957).

\bibitem{Cybenko}
G.~Cybenko,
Approximations by Superpositions of Sigmoidal Functions,
Mathematics of Control, Signals, and Systems, 2(4), pp. 303–314, (1989).

\bibitem{Hornik}
K.~Hornik,
Approximation Capabilities of Multilayer Feedforward Networks,
Neural Networks, 4(2), pp. 251–257, (1991).

\bibitem{eigen}
J.~Demmel,
Computing Small Singular Values of Bidiagonal Matrices with Guaranteed High Relative Accuracy,
Forgotten Books, (2018).

\bibitem{lapack}
E.~Anderson, Z.~Bai, C.~Bischof, S.~Blackford, J.~Demmel, J.~Dongarra, J.~Du~Croz, A.~Greenbaum, S.~Hammarling, A.~McKenney, D.~Sorensen,
LAPACK Users Guide,
Society for Industrial and Applied Mathematics, (1999).

\bibitem{Schroedinger}
E.~Schr\"{o}dinger,
Quantisierung als Eigenwertproblem,
Ann. Phys. 385, pp. 437–490, (1926).
	
\bibitem{Keldysh}
L.V.~Keldysh,
Zh. Eksp. Teor. Fiz.,
Sov. Phys. JETP 20, (1965).

\bibitem{Feynman}
R.P.~Feynman,
Space–time Approach to Non-relativistic Quantum Mechanics,
Rev. Modern Phys. 20, p. 367, (1948).

\bibitem{Wigner}
E.~Wigner,
On the Quantum Correction for Thermodynamic Equilibrium,
Physical Review 40, no. 5, 749, (1932).

\bibitem{SPF}
J.M.~Sellier,
A Signed Particle Formulation of Non-Relativistic Quantum Mechanics,
Journal of Computational Physics 297, pp. 254-265, (2015).

\bibitem{Bishop}
C.M.~Bishop,
Neural Networks for Pattern Recognition,
Oxford University Press, (1995).

\bibitem{Bengio}
I.~Goodfellow, Y.~Bengio, A.~Courville,
Deep Learning,
The MIT Press, (2016).

\bibitem{CMA}
N.~Hansen,
The CMA Evolution Strategy: a Comparing Review,
Towards a New Evolutionary Computation. Studies in Fuzziness and Soft Computing, vol 192, Springer, (2006).

\end{thebibliography}
\end{document}